\documentclass[fleqn,twoside]{article}
\usepackage{espcrc2}

\usepackage{graphicx}
\usepackage[figuresright]{rotating}

\newcommand{\AmS}{{\protect\the\textfont2
  A\kern-.1667em\lower.5ex\hbox{M}\kern-.125emS}}

\def\lesssim{\mathrel{\hbox{\rlap{\hbox{\lower4pt\hbox{$\sim$}}}\hbox{$<$}}}}
\def\gtrsim{\mathrel{\hbox{\rlap{\hbox{\lower4pt\hbox{$\sim$}}}\hbox{$>$}}}}

\title{Primordial bound systems of superheavy
particles\\ as the source of ultra-high energy cosmic rays}

\author{V.K.Dubrovich \address[SAO]
    {Special Astrophysical Observatory RAS, \\
    196140 St.Petersburg, Russia}%
    D.Fargion \address[Uniroma1]
        {Physics department, Universita' degli studi "La Sapienza", \\
         Piazzale Aldo Moro 5,CAP 00185 Roma, Italy and \\
        INFN Roma, Istituto Nazionale di Fisica Nucleare, Italy}%
        and M.Yu.Khlopov
       \address[KHLO]
        {Centre for Cosmoparticle physics "Cosmion", \\
        Miusskaya Pl. 4, 125047, Moscow, Russia;  \\
       Keldysh Institute of Applied Mathematics,  \\
      Miusskaya Pl. 4, 125047, Moscow, Russia;  \\
        Physics department, Universita' degli studi "La
        Sapienza"  and \\
       Moscow Engineering Physics Institute, MEPI , Moscow, Russia}%
}

\begin{document}

\begin{abstract}
Annihilation of superheavy particles in primordial bound systems is considered as the source of
Ultra High Energy Cosmic Rays (UHECR).
Charge conservation makes them to be produced in pairs,
and the estimated separation of particle and antiparticle in such pair
is shown to be in some cases much smaller than the average separation determined
by the averaged number density of considered particles.
If the new U(1) charge is the source of a long range field similar to
electromagnetic field, the particle and antiparticle, possessing that charge, can form primordial bound
system with annihilation timescale, which can satisfy the conditions,
assumed for this type of UHECR sources. These conditions severely constrain the possible properties of considered particles.
So, the proposed mechanism of UHECR origin is impossible to realise, if the U(1) charged particles
share ordinary weak, strong or electromagnetic interactions.
It makes the proposed mechanism of pairing and binding of superheavy U(1) charged particles
an effective theoretical tool in the probes of the physics of very early Universe and of the hidden
sector of particle theory, underlying it.

\vspace{1pc}
\end{abstract}

\maketitle

\section{INTRODUCTION}

Ultra High Energy Cosmic Rays (UHECR) is the observed effect of
superhigh energy physics in the modern Universe, what naturally
puts together particle physics and cosmology in the analysis of
their possible origin and effects. Particle theory predicts a
wide variety of new phenomena in this energy range. Such
phenomena  are unavoidable in the modern Big bang cosmology,
based on inflationary models with baryosynthesis and
(multicomponent?) nonbaryonic dark matter. The physics of
inflation and baryosynthesis, as well as dark matter and energy
content implies new particles, fields and mechanisms, predicted
in the hidden sector of particle theory. Such particles, fields
and mechanisms may play an important role in the problem of
UHECR. It makes new physics necessary component of the analysis
of UHECR data.

The origin of cosmic rays with energies, exceeding the GZK cut off
energy \cite{Greis}, is widely discussed as the possible effect of new
physics. One of popular approaches is related with decays or
annihilation in the Galaxy of primordial super-heavy particles \cite{Berez1,Kuzmin}
(see \cite{Ziaee} for review and references where in).

The mass of such super-heavy  particles to be considered in
present paper is assumed to be higher than the re-heating
temperature of inflationary Universe, so it is assumed that the
particles are created in some non-equilibrium processes (see e.g.
\cite{Chung,Kofman}), taking place after inflation at the stage of preheating.

The problems, related with this approach, are as follows. If the source of
ultra high energy cosmic rays (UHECR) is related with particle decay in
the Galaxy, the timescale of this decay, which is necessary
to reproduce the UHECR data, needs special nontrivial explanation. Indeed,
the relic unstable particle should survive to the present time, and having
the mass $m$ of the order of $10^{14}$ GeV or larger it should have the
lifetime $\tau$,
exceeding the age of the Universe. On the other hand, even, if particle decay
is due to gravitational interaction, and its probability is of the order of
(here and further, if not directly indicated otherwise,
we use the units $\hbar = c = k = 1$)
\begin{equation}
\frac{1}{\tau}=(\frac{m}{M_{P}})^{4} m,
\label{Pltau}
\end{equation}
where $M_{P}=10^{19}$ GeV is the Planck mass, the estimated
lifetime would be by many orders of magnitude smaller. It implies
some specific additional suppression factors in the probability
of decay, which need rather nontrivial physical realization
(\cite{Berez1,Ziaee}), e.g. in the model of cryptons \cite{Ellis} (see \cite{Sarkar} for
review).

If the considered particles are absolutely stable,  the source
of UHECRs is related with their annihilation in the Galaxy. But their
averaged number density, constrained by the upper limit on their total density,
is so low, that strongly inhomogeneous distribution is needed to enhance the
effect of annihilation to the level, desired to explain the origin of UHECR by
this mechanism.

In the present paper, we consider in more details new approach to
the solution of the latter problem, offered in \cite{DubrKhl}. If
superheavy particles possess new U(1) gauge charge, related to
the hidden sector of particle theory, they are created in pairs.
The  Coulomb-like attraction (mediated by the massless U(1) gauge
boson) between particles and antiparticles in these pairs can
lead to their primordial binding, so that the annihilation in the
bound system provides the mechanism for UHECR origin. To realize
this mechanism the properties of superheavy particles should
satisfy a set of conditions, putting severe constrains on the
cosmological scenarios and particle models, underlying the
proposed mechanism.

%%%%%%%%%%%%%%%%%%%%%%%%%%%%%%%%%%%%%%%%%%%%%%%%%%%%%%%%%%%%%%%%%%%%%%%%%%%%%%%%

The necessary decoupling of superheavy particles from the interactions
of ordinary particles can be related with physics of neutrino mass,
resulting in the dominant annihilation channels to neutrino.
It may be important for another approach to a possible solution of the
GZK paradox that considers the Ultra High Energy Cosmic
Rays as secondary products of UHE neutrinos, originated at far
cosmic distances, overcoming GZK cut-off, hitting onto relic light
neutrino in Hot Dark Halos, leading to resonant Z boson
production.  A consequent Z-Shower (Z-Burst)(see \cite{Fargi,Weiler,YoshidaSL,Fargion01,Fodor}) takes place,
where a boosted ultra-relativistic gauge boson Z (or WW, ZZ pairs)
decays in flight and where its UHE nuclear secondaries are the
observed UHECR events in terrestrial atmosphere. These ZeV
primary UHE neutrinos may be produced either inside compact
astrophysical objects (Jets GRBs, AGNs,BL Lac
\cite{Tinyakov-Tkachev2001} ) or by relic topological defects
decay \cite{Kalashev:2002kx} or, as in the present paper, by
superheavy particle annihilation in primordial bound systems. In the first (compact object)
case one may easily understand the observed UHECR clustering  as
well as the possible correlation found recently with BL Lac
sources. In the second case one may assume the UHECR clustering
toward BL Lac  as a pure coincidence; otherwise, one may consider
 a possible faster induced annihilation of these bound systems
 inside deeper clustered gravitational wells around AGN and BL Lacs objects,
treating such annihilation as the source of UHE neutrinos.
%%%%%%%%%%%%%%%%%%%%%%%%%%%%%%%%%%%%%%%%%%%%%%%%%%%%%%%%%%%%%%%%%%%%%%%%%%%%%%%%%%%%%%%%%%%%%%

 \section{Non-equilibrium mechanisms of superheavy particles production}
The modern cosmological models of inflationary Universe assume
that thermodynamically equilibrium conditions of hot Universe
(the so called "reheating") do not take place immediately after
the end of inflation, and that there exist rather long transitions
period of the so called "preheating". The non-equilibrium
character of superheavy particle production implies strong
dependence on the concrete physical processes that can take place
at different periods of preheating stage.

It was shown in \cite{Chung} that the parametric resonance \cite{Kofman} in the end of
inflation at $t \sim 1/H_{end}$, when preheating begins, can lead to intensive inflaton field
decay, in which superheavy particles with the mass $m \le 10 H_{end}$
can be produced. Here $H_{end} \sim 10^{13}$GeV is the Hubble constant in the end of inflation.
The calculations \cite{Chung} of primordial concentration of such superheavy particles
exhibit strong dependence on $m/H$ and correspond to a wide range of their
modern densities up to $\Omega_{X} \sim 0.3$.

Superheavy particles can be created in the end of preheating, when
reheating takes place at $t \sim 1/H_{r}$
($H_{r}$ being the Hubble constant in the period of reheating),
if the quanta of inflaton
field contain these particles among the products of decay. The modern density of superheavy
particles is then given by
\begin{equation}
\Omega_{X} = \frac{T_{r}}{T_{RD}} \frac{2m}{m_{\phi}} Br(X),
\end{equation}
where $T_{r} \sim (H_{r} M_{P})^{1/2}$ is the reheating temperature, $T_{RD} \sim 10$eV is
the temperature in the end of radiation dominance stage and in the beginning of the
modern matter dominated stage, $m_{\phi} > 2m$ is the mass of the inflaton field quantum
and $Br(X)$ is the branching ratio of superheavy particles production in inflaton decay.
The condition $\Omega_{X} \le 0.3$ constrains the branching ratio as
\begin{equation}
Br(X) \le 0.1 \frac{T_{RD}}{T_{r}} \frac{m_{\phi}}{m}.
\end{equation}

If inflation ends by the first order phase transition, bubble wall collisions
in the course of true vacuum bubble nucleation can lead to formation of
primordial black holes (PBH) with the mass $M \sim M_{P} \frac{M_{P}}{H_{end}}$
(see \cite{Khl1} for review).
Successive evaporation of such black holes at
\begin{equation}
H_{ev} \sim 1/t_{ev} \sim \frac{M_{P}^{4}}{M^{3}} \sim H_{end}(\frac{H_{end}}{M_{P}})^{2}
\end{equation}
is the source of superheavy particles, when the temperature of PBH evaporation,
increasing with the loss of mass as
$T_{PBH} \sim M_{P}^{2}/M$, reaches $m$. If $\alpha_{X}$ is the fraction of PBH mass,
evaporated in the form of considered superheavy particles, the relationship between the
probability of PBH formation $w$ and $\Omega_{X}$ is given by
\begin{equation}
\Omega_{X} = \frac{T_{r}}{T_{RD}} \alpha_{X} w,
\label{PBHMD}
\end{equation}
for dust-like (MD) expansion law at preheating stage and
\begin{equation}
\Omega_{X} = \frac{(M_{P}H_{end})^{1/2}}{T_{RD}} \alpha_{X} w,
\label{PBHRD}
\end{equation}
for relativistic (RD) expansion at the stage of preheating. In the latter case
corresponding to the Eq.(\ref{PBHRD}), the condition $\Omega_{X} \le 0.3$ leads to
$w \le 3 \cdot 10^{-25}/ \alpha_{X}$. Creation of mini black holes with such a
low probability does not imply first order phase transition after inflation, but it
is possible even from Gaussian "tails" (see \cite{Khl1} for review) of nearly flat ultraviolet spectra,
that are strongly disfavored but
still not excluded within the uncertainty of the recent WMAP measurements
of CMB anisotropy \cite{WMAP} .

The presence of additional dynamically subdominant fields at the inflationary stage
can strongly modify at the small scales the simple picture of nearly flat power spectrum of density fluctuations. It also
leads to the possibility of superheavy particle production in the decay of quanta of such
field, $\phi$, at the preheating stage.
The relationship between $\Omega_{X}$ and the relative
contribution $r$ of the field, $\rho_{\phi}$, into the total density $\rho_{tot}$,
$r = \rho_{\phi}/\rho_{tot}$ in the period of decay, at $\tau \sim 1/H_{d}$,
is given by Eqs. (\ref{PBHMD})-(\ref{PBHRD}), in which $\alpha_{X}$ has the meaning of the branching ratio
for superheavy particle production (multiplied by the factor $\sim m/m_{\phi}$,
$m_{\phi}$ is the mass of $\phi$, in case of relativistic decay products) and
$H_{end}$ is substituted by $H_{d}$. If $\phi$ decays due to gravitational interaction,
$H_{d}$ is equal to the probability of decay, $\Gamma$, given by Eq.(\ref{Pltau}),
$H_{d}= \Gamma \sim m_{\phi}(m_{\phi}/M_{P})^{4}$. In general, for $H_{d}= \Gamma$,
the period of derelativization of the relativistic decay products with the energy
$\epsilon \sim m_{\phi} \gg m$ corresponds to
$H \sim (m/m_{\phi})^{2} H_{d}$.

\section{Primordial pairing of superheavy particles}

Note, first of all, following \cite{DubrKhl}, that in quantum theory particle stability reflects
the conservation law, which according to Noether's theorem is related
with the existence of a conserved charge, possessed by the considered
particle. Charge conservation implies that particle should be created
together with its antiparticle. It means that, being stable,
the considered superheavy
particles should bear a conserved charge, and
such charged particles should be created in pairs with their antiparticles
at the stage of preheating.

Being created in the local process
the pair is localized within the cosmological
horizon in the period of creation. If the momentum distribution
of created particles is peaked below $p \sim mc$, they don't
spread beyond the proper region of their original localization,
being in the period of creation $l \sim c/H$, where
the Hubble constant $H$ at the preheating stage
is in the range $H_{r} \le H \le H_{end}$.
For relativistic pairs the region of localization is determined
by the size of cosmological horizon in the period
of their derelativization.
In the course of successive expansion the distance $l$ between
particles and antiparticles grows with the scale factor,
so that after reheating at the temperature $T$ it is
equal to
\begin{equation}
l(T) = (\frac{M_{P}}{H})^{1/2} \frac{1}{T}.
\label{lsep}
\end{equation}

The averaged number density of superheavy particles $n$
is constrained by the upper limit on their modern density.
Say, if we take their maximal possible contribution in the units of critical density, $\Omega_{X}$, not to exceed
0.3, the modern cosmological average number
density should be $n = 10^{-20}\frac{10^{14} GeV}{m} \frac{\Omega_{X}}{0.3}cm^{-3}$
(being $n= 4 \cdot 10^{-22} \frac{10^{14}GeV}{m} \frac{\Omega_{X}}{0.3} T^{3}$ in
 the units $\hbar = c = k = 1$ at the temperature $T$). It correponds
at the temperature $T$ (at the redshift $z$) to the mean distance ($l_{s} \sim n^{-1/3}$)
equal to
$$l_{s} \approx 1.6 \cdot 10^{7} (\frac{m}{10^{14}GeV})^{1/3}(\frac{0.3}{\Omega_{X}})^{1/3} \frac{1}{T} \approx $$
\begin{equation}
\approx 4.6 \cdot 10^{6}(\frac{m}{10^{14}GeV})^{1/3}(\frac{0.3}{\Omega_{X}})^{1/3} (1+z)^{-1} cm.
\label{lmean}
\end{equation}

One finds that superheavy nonrelativistic particles,
created just after the end of inflation,
when $H \sim H_{end} \sim 10^{13} GeV$, are separated from their
antiparticles at distances more than 4 orders of magnitude
smaller, than the average distance between these pairs. On the other hand,
if the nonequilibrium processes of superheavy particles creation
(such as decay of inflaton) take place in the end of preheating stage,
and the reheating temperature is as low as it is constrained from the effects
of gravitino decays on $^{6}Li$ abundance ($T_{r}< 4\cdot 10^{6}$GeV \cite{Khl1,Khl2}),
the primordial separation of pairs, given by Eq(\ref{lsep}), can even exceed the value, given by Eq.(\ref{lmean}).
It means that the separation between particles and antiparticles can be determined
in this case by their
averaged density, if they were created at $H \le H_{s}\sim 10^{-15} \cdot M_{P}
(\frac{10^{14}GeV}{m})^{2/3}(\frac{\Omega_{X}}{0.3})^{2/3}
\sim 10^{4}(\frac{\Omega_{X}}{0.3})^{2/3}(\frac{10^{14}GeV}{m})^{2/3}$GeV.

If the considered charge is the source of a long range field,
similar to the electromagnetic field, which can bind
particle and antiparticle into the atom-like system,
analogous to positronium, it may have important practical
implications for UHECR problem. The annihilation timescale
of such bound system can provide the rate of UHE particle sources,
corresponding to UHECR data.

\section{Formation of primordial bound systems from primordial pairs}

The pair of particle and antiparticle with opposite gauge charges
forms bound system, when in the course of expansion the absolute
magnitude of potential energy of pair $V= \frac{\alpha_{y}}{l}$
exceeds the kinetic energy of particle relative motion $T_{k}=
\frac{p^{2}}{2m}$. The mechanism is similar to the proposed in
\cite{Dubr} for binding of magnetic monopole-antimonopole pairs. It is
not a recombination one. The binding of two opposite charged
particles is caused just by their Coulomb-like attraction, once
it exceeds the kinetic energy of their relative motion.

In case, plasma interactions do not heat superheavy particles,
created with relative momentum $p \le mc$ in the period,
corresponding to Hubble constant $H \ge H_{s}$, their initial
separation, being of the order of
\begin{equation}
l(H) = (\frac{p}{mH}),
\label{lins}
\end{equation}
experiences only the effect of general expansion, proportional to the inverse
first power of the scale factor, while the initial kinetic energy decreases
as the square of the scale factor. Thus, the binding condition is fulfilled
in the period, corresponding to the Hubble constant $H_{c}$, determined by
the equation
\begin{equation}
(\frac{H}{H_{c}})^{1/2} = \frac{p^{3}}{2 m^{2}\alpha_{y}H},
\label{bindcond}
\end{equation}
where $H$ is the Hubble constant in the period of particle creation and $\alpha_{y}$ is the "running constant" of the long range interaction,
possessed by the superheavy particles. If the local process of pair creation does not involve nonzero orbital momentum, due to the
primordial pairing the bound system is formed in the state with zero orbital momentum.
The size of bound system
exhibits strong dependence on the initial momentum distribution
$$l_{c} = \frac{p^{4}}{2 \alpha_{y} m^{3} H^{2}}= 2 \frac{\alpha_{y}}{m \beta^{2}} \approx$$
\begin{equation}
\approx 8 \cdot 10^{-6} (50 \alpha_{y})(\frac{10^{14}GeV}{m})(\frac{10^{-12}}{\beta})^{2} cm,
\label{lbind}
\end{equation}
where
\begin{equation}
\beta = \frac{2 \alpha_{y} m H}{p^{2}}.
\end{equation}
In principle, it facilitates the possibility to fit UHECR data
in the framework of hypothesis of bound system annihilation
in the halo of our Galaxy.

Indeed, the annihilation timescale of this bound system can
be estimated from the annihilation rate, given by
\begin{equation}
w_{ann} = \mid \Psi (0) \mid ^{2} (\sigma v)_{ann} \sim l_{c}^{-3}
\frac{\alpha_{y} ^{2}}{m^{2}} C_{y},
\label{wannc}
\end{equation}
where the "Coulomb" factor $C_{y}$ arises similar to
the case of a pair of electrically charged particle and antiparticle.
For the relative velocity $\frac{v}{c}\ll 1$ it is given by \cite{Khlopov}
\begin{equation}
C_{y} = \frac{2 \pi \alpha_{y} c}{v}.
\end{equation}
Finally, taking $v/c \sim \beta$, one obtains for the annihilation timescale
\begin{equation}
\tau_{ann} \sim \frac{1}{8 \pi \alpha_{y}^{5}} (\frac{p}{mc})^{10}
(\frac{m}{H})^{5} \frac{1}{m} = \frac{4}{\pi m \beta^{5}}.
\label{tausep}
\end{equation}

For $H_{end}\ge H \ge H_{s}$ , the
annihilation timescale equals
\begin{equation}
\tau_{ann} = 10^{22} (\frac{10^{14}GeV}{m})(\frac{10^{-12}}{\beta})^{5} s,
\label{tauan}
\end{equation}
being for $p \sim mc$, $\alpha_{y}= \frac{1}{50}$
and $m = 10^{14}$ GeV in the range from $10^{-26}$s
up to $10^{19}(\frac{0.3}{\Omega_{X}})^{10/3}$s.
The size of a bound system is given by
\begin{equation}
l_{c} = 8 \cdot 10^{-6}  (50\alpha_{y}) (\frac{10^{14}GeV}{m})
(\frac{10^{-12}}{\beta})^{2} cm,
\label{lsys}
\end{equation}
ranging for $2 \cdot 10^{-10} \le \frac{\Omega_{X}}{0.3} \le 1$
from $7 \cdot 10^{-7}$cm to $6 \cdot 10^{-3}$cm. One can obtain from
Eqs. (\ref{tauan})-(\ref{lsys}) the approximate relationship between $\tau_{ann}$ and $l_{c}$,
given by
\begin{equation}
\frac{\tau_{ann}}{10^{10}yr} \simeq  (\frac{l_{c}}{10^{-7}cm})^{5/2}.
\end{equation}

Provided that the primordial abundance of superheavy particles,
created on preheating stage corresponds to the appropriate modern density
$\Omega_{X} \le 0.3$, and the annihilation timescale exceeds the age of the Universe
$t_{U} = 4 \cdot 10^{17}$s, owing to strong dependence on the parameter $\beta$, the magnitude
\begin{equation}
r_{X} = \frac{\Omega_{X}}{0.3} \frac{t_{U}}{\tau_{X}}
\label{rx}
\end{equation}
can easily take the value $r_{X} = 2 \cdot 10^{-10}$,
which was found in \cite{Berez1} to fit the UHECR data by superheavy
particle decays in the halo of our Galaxy. It takes place,
provided that
\begin{equation}
(\frac{\Omega_{X}}{0.3})(\frac{m}{10^{14}GeV})(\frac{\beta}{10^{-12}})^{5} = 5 \cdot 10^{-6}.
\label{fit}
\end{equation}

If the effective production of superheavy particles takes place in the end of
preheating stage at $H \le H_{s}$, their initial separation
is determined by the $\min\{l(H),l_{s}\}$, where $l(H)$ is given by Eq.(\ref{lins}) and $l_{s}$ is determined by
their mean number density (compare with Eq.(\ref{lmean}))
$$l_{s} \approx 3 \cdot 10^{7} (\frac{m}{10^{14}GeV})^{1/3} (\frac{0.3}{\Omega_{X}})^{1/3} (\frac{1}{M_{P}H})^{1/2} \approx$$
\begin{equation}
\approx 10^{-4}(\frac{m}{10^{14}GeV})^{1/3} (\frac{0.3}{\Omega_{X}})^{1/3}(\frac{10^{4}GeV}{H})^{1/2} cm.
\label{lmeans}
\end{equation}

In the case of late particle production (i.e. at $H \le H_{s}$) the binding condition
can retain the form (\ref{bindcond}), if $l(H) \le l_{s}$.
Then the previous estimations (\ref{lbind})-(\ref{tausep}) are valid.

\section{Formation of primordial bound systems without primordial pairing}
In the opposite case  of late particle production, when $l(H) \ge l_{s}$,
the primordial pairing is lost and even being produced
with zero orbital momentum particles and antiparticles, originated from different pairs,
in general, form bound systems with nonzero orbital momentum.
The size of the bound system is in this case obtained from the binding condition
for the initial separation, determined by Eq.(\ref{lmeans}), and it is equal to
$$l_{c} \approx \frac{10^{15}}{2 \alpha_{y} M_{P}}
(\frac{m}{10^{14}GeV})^{2/3} (\frac{0.3}{\Omega_{X}})^{2/3}
(\frac{p}{mc})^{2}(\frac{m}{H})$$
\begin{equation}
\approx \frac{1}{H_{s}} \approx 2 \cdot 10^{-6}(\frac{m}{10^{14}GeV})^{2/3}
(\frac{0.3}{\Omega_{X}})^{2/3}(\frac{10^{-12}}{\beta})cm.
\label{llate}
\end{equation}
The orbital momentum of this bound system can be estimated as $M
\sim mvl_{c}$ and the lifetime of such bound system is determined
by the timescale of the loss of this orbital momentum. This
timescale can be reasonably estimated with the use of the well
known results of classical problem of the falling down the center
due to radiation in the bound system of opposite electric charges
$e_{1}$ and $e_{2}$ with masses $m_{1}$ and $m_{2}$, initial
orbital momentum $M$ and absolute value of the initial binding
energy $E$ (see e.g. \cite{Landau})
$$t_{f} = \frac{c^{3}M^{5}}{\alpha_{y} (2 E
\mu^{3})^{1/2}}(\frac{e_{1}}{m_{1}} -
\frac{e_{2}}{m_{2}})^{2}\cdot$$
\begin{equation}((\mu \alpha_{y}^{2})^{1/2} + (2
M^{2} E)^{1/2})^{-2}.
\label{trad}
\end{equation}
Here $\mu= \frac{m_{1}m_{2}}{m_{1}+ m_{2}}$ is the reduced mass. Putting into Eq.(\ref{trad})
$M= \mu v l_{c}$, $E= \mu v^{2}/2$, and with the account for $2 M^{2} E \sim \mu \alpha_{y}^{2}$
one obtains the lifetime of the bound system as
$$\tau =  \frac{l_{c}^{3}}{64 \pi} \frac{m^{2}}{\alpha_{y}^{2}}=$$
\begin{equation}
= 4 \cdot 10^{20}(\frac{l_{c}}{10^{-6}cm})^{3}(\frac{m}{10^{14}GeV})^{2}(50\alpha_{y})^{-2}yr.
\label{taulate}
\end{equation}
Using the Eq.(\ref{wannc}) and the condition $l(H) \ge l_{s}$, one obtains for this case
the following restriction
\begin{equation}
r_{X} = \frac{\Omega_{X}}{0.3} \frac{t_{U}}{\tau_{X}} \le 3 \cdot 10^{-10} (\frac{\Omega_{X}}{0.3})^{5} (\frac{10^{14}GeV}{m})^{9}.
\label{restr}
\end{equation}
Note that the condition (\ref{restr}) admits $\Omega_{X}= 0.3$, when superheavy particles dominate in
the dark matter of the modern Universe.

The gauge U(1) nature of the charge, possessed by superheavy particles, assumes the existence of massless U(1) gauge bosons
(y-photons) mediating this interaction. Since the considered superheavy particles are the lightest particles
bearing this charge, and they are not in thermodynamical equilibrium, one can expect that there should be no thermal background
of y-photons and that their non equilibrium fluxes can not heat significantly the superheavy particles.

\section{Primordial bound systems in the case of plasma heating}
The situation changes drastically, if the superheavy particles  possess not only new U(1) charge but also
some ordinary (weak, strong or electric) charge \cite{DubrKhl}.
Due to this charge superheavy particles interact with the equilibrium relativistic plasma
(with the number density $n \sim T^{3}$)
and for the mass of particles $m \le \alpha^{2} M_{P}$ the rate of heating
\begin{equation}
n \sigma v \Delta E \sim \alpha^{2} \frac{T^{3}}{m}
\end{equation}
is sufficiently high
to bring the particles into thermal
equilibrium with this plasma.
Here $\alpha$ is the running constant of
the considered (weak, strong or electromagnetic) interaction.

Plasma heating
causes the thermal motion of superheavy particles.
At $T \le m (\frac{m}{\alpha^{2} M_{P}})^{2}$ their mean free path
relative to scattering with plasma exceeds the free thermal motion path,
so it is not diffusion, but free motion with thermal velocity $v_{T}$ that leads to
complete loss of initial pairing, since  $v_{T}t$ formally exceeds $l_{s}$
at $T \le 10^{-10} M_{P}(\frac{\Omega_{X}}{0.3})^{2/3} (\frac{10^{14}GeV}{m})^{5/3}$.

In the case, the interaction with plasma keeps superheavy
particles in thermal equilibrium, potential energy of charge
interaction $V = \frac{\alpha_{y}}{l_{s}}$ is less, than thermal
energy $T$ for any $\alpha_{y} \le 3 \cdot 10^{7}
(\frac{0.3}{\Omega_{X}})^{1/3} (\frac{m}{10^{14}GeV})^{1/3}$. So
binding condition $V \ge T_{kin}$ can not take place, when plasma
heating of superheavy particles is effective.

For electrically charged particles it is the case until
electron positron pairs annihilate at $T_{e} \sim 100$keV (see \cite{Dubr})
and for colored particles until QCD phase transition at $T_{QCD} \sim 300$MeV.
In the latter case colored superheavy particles form superheavy stable hadrons,
possessing U(1) charge.
For weakly interacting particles after electroweak phase transition,
when Eq. (\ref{fit}) is not valid,
neutrino heating, given by $n \sigma v \Delta E \sim G_{F}^{2} \frac{T^{7}}{m}$,
 is sufficiently effective until $T_{w} \approx 20$GeV. At $T < T_{N}$, where
$N = e, QCD, w$ respectively, the plasma heating is suppressed
and superheavy particles go out of thermal equilibrium.

In the course of successive expansion kinetic energy of superheavy
particles
falls down with the scale factor $a$ as $\propto a^{2}$, and the binding
condition is reached at $T_{c}$, given by
\begin{equation}
T_{c} = T_{N} \alpha_{y} 3 \cdot 10^{-8} (\frac{\Omega_{X}}{0.3})^{1/3} (\frac{10^{14}GeV}{m})^{1/3}.
\label{Tcrad}
\end{equation}
For electrically charged particles, forming after recombination atom-like states with protons and
electrons, but still experiencing the Coulomb-like attraction due to non-compensated U(1) charges,
the binding in fact does not take place to the present time, since  one gets from Eq.(\ref{Tcrad}) $T_{c} \le 1$K.
Bound systems of hadronic and weakly interacting superheavy particles can form, respectively,
at $T_{c} \sim 0.3$eV and $T_{c} \approx 20$eV.

The size of the bound system is then given by
\begin{equation}
l_{c} = 10^{15} (\frac{0.3}{\Omega_{X}})^{2/3} (\frac{m}{10^{14}GeV})^{2/3} \frac{1}{\alpha_{y} T_{N}} \approx
\end{equation}
$$\approx 10^{3} (\frac{0.3}{\Omega_{X}})^{2/3} (\frac{m}{10^{14}GeV})^{2/3} (50 \alpha_{y})^{-1} \frac{1 GeV}{T_{N}}cm,$$
what even for weakly interacting particles approaches a half of meter (30 m for hadronic particles!). It leads to extremely long annihilation timescale of
these bound systems, that can not fit UHECR data. Moreover, being extremely weakly bound, they should be disrupted
almost completely, colliding in Galaxy. So, for bound systems of weakly interacting superheavy particles, $n \sigma v t_{U} \sim 10^{14}$
where $n = 3 \cdot 10^{-15} cm^{-3} \frac{10^{14} GeV}{m} \frac{\Omega_{X}}{0.3}$
is the number density of bound systems, $\sigma \sim \pi l_{c}^{2}$ and their
relative velocity $v \sim 3 \cdot 10^{7}$cm/s.
It makes impossible to realise the considered mechanism of UHECR origin, if the superheavy U(1) charged particles share
ordinary weak, strong or electromagnetic interactions.
\section{Evolution of bound systems in the Galaxy}

Superheavy particles, as any
other form of CDM should participate gravitational clustering
and concentrate to the center in the course of Galaxy formation.
There are several factors influencing the evolution of
bound systems in the Galaxy.

If the size of primordial bound systems
$l_{c} \ge 3 \cdot 10^{-6} cm (\frac{0.3}{\Omega_{X}} \frac{m}{10^{14}GeV})^{1/2}$
their collision rate in the vicinity of Solar system
$n \sigma v t_{U} \ge 1$. It can take place,
provided that
\begin{equation}
(50 \alpha_{y})^{4} (\frac{10^{14}GeV}{m})^{11/5} (\frac{\Omega_{X}}{0.3})^{9/5} >  6 \cdot 10^{-6}.
\label{coll}
\end{equation}

Since the binding energy of bound systems
($E_{b} \le 5 \cdot 10^{-27} mc^{2} (\frac{0.3}{\Omega_{X}} \frac{10^{14}GeV}{m})^{2/5}$) is much less
than the kinetic energy of their motion in the Galaxy ($T = m v^{2}/2$), the bound systems
should be disrupted in such collisions. On the other hand,
large collision cross section $\sigma \sim \pi l_{c}^{2}$
corresponds to the momentum transfer of the order of $\Delta p \sim m \beta c$.
Such small momentum transfer leads both to
disruption of the bound system on free superheavy particles
and with the same order of probability to the reduction of their size down to
$l \sim l_{c}/4$. If the bound systems are within gravitationally
bound cluster (Galaxy, globular cluster, CDM small scale cluster),
both the free particles and contracted bound systems
with the size $l$ remain in it and the collisions both between
bound systems and between free particles and bound systems continue
with the cross section $\sigma \sim \pi l^{2}$.

In vicinity of massive objects with the mass $M$ tidal effects
lead to disruption of bound systems with the size $l$ at the
distance $r$, corresponding to tidal force energy $\sim
\frac{GMm}{r} \frac{l}{r}$, exceeding the binding energy $\sim
\alpha_{y}/l$. So, in the vicinity of a star with the mass $M =
M_{\odot}$ the bound systems are disrupted at the distances,
smaller than $r_{d}= \sqrt{\frac{GMm}{\alpha_{y}}}l$. Tidal
effects disrupt bound systems in the regions of enhanced stellar
density, $n_{M}$, if $n_{M} \pi r_{d}^{2} v t_{U} \ge 1$. Taking
for the center of Galaxy $n_{M} \sim 10^{6} M_{\odot}/pc^{3}$,
one finds that bound systems with the size $l \ge 5 \cdot
10^{-6}$cm should be disrupted there due to tidal effects.

As it was recently shown in \cite{Dok}, tidal effects strongly influence
the formation and mass distribution of small scale CDM clusters,
what should also take place for small scale clusters
of primordial bound systems in the Galaxy. On the other hand,
clustering of primordial bound systems may play important
role in the explanation of observed clustering of UHECR events
in the framework of the proposed mechanism. It may be easily estimated that
if bound systems are clustered around globular cluster,
their disruption due to stellar tidal effects is negligible.

\section{Space distribution and clustering of UHECR events}
For the initial number density and size of bound systems,
corresponding to $n \sigma v t_{U} \ge 1$, most of bound systems
disrupt on the free particles, but sufficiently large fraction
of them $\sim \beta_{c}/\beta_{U} \sim (l_{U}/l_{c})^{1/2}$
acquires the size $l_{U}$, at which $n \sigma v t_{U} \approx 1$. The relative amount of bound systems
with smaller size $l < l_{U}$ is of the order of $\sim (l/l_{U})^{2}$,
if their annihilation timescale $\tau \ge t_{U}$ and
of the order of $\sim (l/l_{U})^{2} \frac{\tau}{t_{U}}$ for
$\tau \le t_{U}$. Annihilation of superheavy particles in
bound systems with the smaller size is more
rapid, what increases the production rate of such UHECR source as
compared with the case of superheavy decaying particles with
the fixed lifetime. This effect of the self-adjustment of
bound systems annihilation leads to the peculiar space
distribution of UHECR sources, corresponding to this mechanism.

The decay rate density, $q$, of metastable particles with lifetime $\tau$
for the number density $n(R)$, depending on the distance $R$ from the center
of Galaxy,
 is given by
\begin{equation}
q = (\frac{n(R)}{\tau}).
\end{equation}
Owing to self adjustment of
bound systems annihilation the density of UHECR production rate
for the initial number density $n(R)$ of primordial bound systems with initial
annihilation timescale $\tau$ has the order of the magnitude
\begin{equation}
q = (\frac{n(R)}{\tau})(\frac{\tau}{\tau_{U}})^{3/5}(\frac{\tau}{t_{U}})^{1/5}.
\label{qbind}
\end{equation}
Since $\tau_{U} \propto l_{U}^{5/2} \propto n(R)^{-5/4}$ the self-adjustment of
bound systems leads to stronger radial dependence $q \propto n(R)^{7/4}$,
thus sharpening the concentration of UHECR sources to the center of Galaxy.

In the case of late particle production with $l(H) \ge l_{s}$, considered in the Section 5,
the dependence $l \propto \beta^{-1}$, given by Eq.(\ref{llate}), and $\tau \propto l^{3} \propto \beta^{-3}$,
given by Eq.(\ref{taulate}), results in the change of Eq.(\ref{qbind}) by
$$q = (\frac{n(R)}{\tau})(\frac{\tau}{\tau_{U}})^{2/3}(\frac{\tau}{t_{U}})^{1/3}.$$
Since $\tau_{U} \propto l_{U}^{3} \propto n(R)^{-3/2}$ in this case, the radial
dependence $q \propto n(R)^{3/2}$ provides the principal possibility
to distinguish this case from the case $l(H) \le l_{s}$.

It was noticed in \cite{Kusen} that clustering of UHECR events, observed in AGASA experiment\cite{AGASA},
can be explained in the model of superheavy metastable particles, if such particles
with the mass $m \sim 10^{14} GeV$
form clusters in the Galaxy with the mass $M \sim 5 \cdot M_{\odot} \frac{\tau_{X}}{10^{10}yr}$.

For the cluster of N metastable particles with lifetime $\tau$
the decay rate, $P$, is given by
\begin{equation}
P = (\frac{N}{\tau}).
\end{equation}
Owing to self adjustment of
bound systems annihilation the UHECR production rate
for the cluster of N primordial bound systems with initial
annihilation timescale $\tau$ reaches  the order of the magnitude
\begin{equation}
P = (\frac{N}{\tau})(\frac{\tau}{\tau_{U}})^{3/5}(\frac{\tau}{t_{U}})^{1/5}.
\end{equation}

The enhancement of UHECR production rate due to self-adjustment of bound systems
facilitates the possibility to explain clustering of UHECR events in the proposed
mechanism. Say, instead of cluster with the mass of $\sim 5 \cdot 10^{6} M_{\odot}$
of particles with the mass $m \sim 10^{14} GeV$ and the lifetime $\tau \sim 10^{16}$yr,
corresponding to the considered in \cite{Kusen}
possible explanation for the clustering of UHECR events, observed by AGASA,
it is sufficient to have the mass of $\sim 3 \cdot 10^{3} M_{\odot}$ in the cluster
of primordial bounds systems with the same initial annihilation timescale
$\tau \approx 10^{16}$yr.

In the case of late particle production with $l(H) \ge l_{s}$, for $p \sim mc$,
the condition $H < H_{s}$ constrains the annihilation timescale
$$\tau > 10^{26}s(\frac{m}{10^{14}GeV})^{9}(50\alpha_{y})^{-3} =$$
$$= 3 \cdot 10^{18}yr(\frac{m}{10^{14}GeV})^{9}(50\alpha_{y})^{-3},$$
the size of the bound system $l$ being related with $\tau$ by Eq.(\ref{taulate}).
In this case, owing to the effect of self-adjustment, clustering of UHECR events can
be also reproduced, e.g. in cluster with the mass $M \sim 10^{5} M_{\odot}$
and number density $n \sim 10^{-10} cm^{-3}$ of superheavy particles with
mass $m \sim 10^{13} GeV$ and initial annihilation timescale of their bound
systems $\tau = 4 \cdot 10^{18}yr$ (for metastable particles
with the same mass and lifetime the cluster with the mass $4 \cdot 10^{8} M_{\odot}$
is needed). Taking into account the possibility of dominance of superheavy
particles in the modern CDM, mentioned above for the considered case,
the estimated parameters of such cluster seem to be reasonable.

Owing to the effect of self-adjustment the intial annihilation time
decreases in the dense regions. On the other hand, disruption of bound systems in collisions
leads to the decrease of their actual amount in the modern Universe as compared with
their primordial abundance. It leads to the corresponding corrections
in the conditions (\ref{fit}) and (\ref{restr}). Provided that the inequality (\ref{coll})
is valid and the collisions of bound systems are significant, the annihilation
timescale $\tau$ should be corrected by the effect of self-adjustment
and one should substitute $\Omega_{X}$ by $\Omega_{bs} < \Omega_{X}$,
where $\Omega_{bs}$ is the averaged concentration of bound systems,
surviving after collisions.

\section{Annihilation into ordinary particles}
To be the source of UHECR the products of superheavy particles annihilation
should contain significant amount of ordinary particles.
On the other hand, it was shown in the present paper that to be the viable source of UHECR
the considered particles should not possess ordinary strong,
weak and electromagnetic interactions. Their interaction with ordinary particles,
giving rise to UHECR production in their annihilation should be related to the
superhighenergy sector of particle theory and/or physics of inflation and preheating.
The selfconsistent treatment of this problem should involve
the realistic particle physics model, reproducing the desired features of inflationary scenario
and giving detailed predictions for physical properties of U(1) charged superheavy particles.
It may be expected that in such models, there can exist superheavy boson (Y),
interacting both with superheavy and ordinary particles. If it's mass is of the order
of $m_{Y} \ge m$ the annihilation channel into ordinary particles with the cross section
$\sigma \sim \frac{\alpha_{Y}^{2}}{m_{W}^{4}} m^{2}$ will be of the order of
the cross section of the two y-photon annihilation channel.

With the account for the invisible $yy$ mode of annihilation, as well as taking into account
effects of self-adjustment and destructions of bound systems in collisions,
one has to re-define the magnitude $r_{X}$, given above by Eq. (\ref{rx}),
as
\begin{equation}
r_{X}^{eff} = B_{o}\frac{\Omega_{bs}}{0.3} \frac{t_{U}}{\tau_{eff}},
\label{rxeff}
\end{equation}
where $B_{o}$ is the branching ratio of annihilation channels to ordinary particles and
$\Omega_{bs} < \Omega_{X}$ is the modern cosmological density of bound systems.
In the Eq. (\ref{rxeff}) $\tau_{eff} = (\tau_{X} \tau_{U}^{3} t_{U})^{1/5}$ and $\tau_{eff} = (\tau_{X} \tau_{U} t_{U})^{1/3}$
for the cases of early and late particle production, respectively.

Note that at $B_{o} \sim 10^{-5}$, when annihilation to ordinary particles is
strongly suppresssed, the case $\Omega_{X} \sim 0.3$ is possible also for
early particle production.

Neutrino channel may strongly dominate in the annihilation to ordinary particles,
so that $B_{o} \approx B_{\nu}$ and the channels to other ordinary particles
are strongly suppressed ($B_{op} \ll 10^{-5}$, where $op = q,l,\gamma,g,W,Z,h$).
Then annihilation of bound systems can not be direct local (galactic) source
of UHECR, but it can provide the source of UHE neutrinos
for the Z-Shower mechanism of UHECR origin.

\section{CONCLUSIONS}
The combination of the constrains on the conditions of particle
creation in the early Universe and on the effective production of
UHECR puts additional constrains on the parameters of the
proposed mechanism, which we plan to consider in the framework of
specific models of particle theory, underlying the scenarios of
very early Universe. The evolution of primordial bound systems
should be also analyzed on the base of such models.
One, however, can make the general conclusion that
the two principal types of bound systems are possible, originated from
(i) "Early particle production", when primordial pairing
essentially determines the formation of bound systems and from
(ii) "Late particle production", when the primordial
pairing is not essential for bound system formation.

In the both cases bound systems can dominate in the modern CDM,
but the conditions for such dominance are different. In the case (i)
annihilation of bound systems with $\Omega_{bs} \sim 0.3$ can reproduce
UHECR events, if the branching ratio for annihilation to ordinary particles
is small ($B_{o} \sim 10^{-5}$), whereas in case (ii) this branching ratio
should maximally approach to 1.

The possibility of bound system disruption in their
collisions in galaxies is specific for the considered mechanism,
making it different from the models of decaying and annihilating
superheavy particles. If effective, such disruption results in a
nontrivial situation, when superheavy particles can
dominate in the modern CDM, while the UHECR sources represent a
sparse subdominant component of bound systems, surviving after disruption.

Self-adjustment of bound systems annihilation in the Galaxy
sharpens their concentration to the center of Galaxy and
increases their UHECR production rate in clusters. It
 provides their difference
from the case of metastable particles.
This property, however, crucially depends on
the probability of contraction of bound systems in the course
of collisions. More detailed analysis of such collisions
is needed to prove this result. If proven
to be really specific for the considered type of UHECR
sources, it will be principally possible to distinguish
 them from other possible  mechanisms \cite{Fargi,Sarkar,Berez2}
in the future AUGER and EUSO experiments.

Pair correlation, considered in the present note, takes place, if the local
process of superheavy particle creation
preserves charge conservation. This condition
has serious grounds in the case of a local U(1) gauge charge,
similar to electric charge, but it may not be the case
for global charge, say, for mechanisms of R-parity nonconservation
due to quantum gravity wormhole effects \cite{Berez1}. The crucial physical condition for the
formation of primordial bound systems of superheavy particles is the existence
of new strictly conserved local U(1) gauge symmetry, ascribed to the hidden sector of
particle theory. Such symmetry can arise in the extended variants of GUT
models (see e.g. \cite{Khl1} for review), in heterotic string phenomenology (see \cite{Khlopov} and references wherein)
and in D-brane phenomenology \cite{Alday}. Note, that in such models the strictly conserved symmetry of
hidden sector can be also SU(2), what leads to a nontrivial mechanism of primordial binding of superheavy
particles due to macroscopic size SU(2) confinement, as it was the case for "thetons" \cite{Okun}.

The proposed mechanism is deeply involved into the details of the
hidden sector of particle theory, what may seem rather artificial
and fine-tuned. However, the necessary combination of conditions
(superheavy stable particles, possessing new strictly conserved
U(1) charge, existence of their superheavy Y-boson interaction
with ordinary particles, nontrivial physics of inflation and
preheating) can be rather naturally realised in the hidden sector
of particle theory. In this aspect, the proposed mechanism offers
the link between the observed UHECRs and the predictions of
particle theory, which can not be tested by any other means and
on which the analysis of primordial pairing and binding can put
severe constrains.

%%%%%%%%%%%%%%%%%%%%%%%%%%%%%%%%%%%%%%%%%%%%%%%%%%%%%%%%
Even so, while  we may agree that the number and sequence of the
assumption of present scenario may sound  artificial and ad hoc,
(maybe at the same level of topological decay lifetime)  we have
taken into account a large number of astrophysical and
cosmological bounds narrowing the parameter window  into a very
severe and fragile regime which may soon survive (or not) future
theoretical self-consistence and experimental test. Indeed, if
HIRES and AGASA data will converge to a GZK cut off with no
spectra extensions to Grand Unified energies or, in a different
scenario,  in case of more evidence for UHECR clustering to BL
Lac sources compatible only to Z-Shower model \cite{Fargi,Weiler}, the model may
 be  considered as an untenable direct local solution of UHECR puzzle.
However, for suppressed channels of annihilation to quarks, charged leptons,
gauge and Higgs bosons, annihilation of superheavy particles in
bound systems can provide the effective sourse of UHE neutrinos,
thus playing important role in the UHECR production by Z-Shower mechanism.
It should be remind that UHE neutrinos of all flavours will be produces
  by such heavy particle annihilation leading to important
  signals in new generation UHE neutrino telescopes based on Horizontal Tau
  Showering (or Earth Skimming Neutrinos) \cite{Fargion
  2000-2002} (see review and some answers to the recent
arguments \cite{Semikoz} against Z-Shower mechanism in \cite{Fargion2003}).

%%%%%%%%%%%%%%%%%%%%%%%%%%%%%%%%%%%%%%%%%%%%%%%%%%%%%%%%%

If viable, the considered mechanism makes UHECR the unique source
of detailed information on the possible properties of the hidden
sector of particle theory and on the physics of very early
Universe
\section{ACKNOWLEDGEMENTS}
The work was partially performed in the framework
of State Contract 40.022.1.1.1106
and supported in part by RFBR grant 02-02-17490
and grant UR.02.01.026. One of us (M.Yu.Kh.) expresses his gratitude to
IHES, LUTH (Observatory Paris-Meudon), ICRA (Pescara Center)
for hospitality and the
authors (V.K.D. and M.Yu.Kh.) are grateful for support of
their visit to College de France
(Paris, April 2002).

% \cite{Scho70,Mazu84,Dimi75,Eato75}.
\end{document}